\documentclass[traditabstract]{aa}

\usepackage{graphicx}
\usepackage{txfonts}
\usepackage{natbib}
\usepackage[colorlinks=true,citecolor=blue]{hyperref}
\usepackage{xspace}
\usepackage{array}
\usepackage[labelformat=simple]{caption,subcaption}
\usepackage{bm}
\usepackage{txfonts}
\usepackage{ulem}

\bibpunct{(}{)}{;}{a}{}{,}

\newcommand{\mic}{$\muup$m\xspace}
\newcommand{\as}{\hbox{$^{\prime\prime}$}\xspace}
\newcommand{\lsd}{\hbox{$\lambda/D$}\xspace}

\usepackage{color}

\begin{document} 

\title{Calibration of residual aberrations in exoplanet imagers with large numbers of degrees of freedom}


\author{
    R.~Pourcelot\inst{\ref{lam},\ref{oca}}     
    \and
    A.~Vigan\inst{\ref{lam}}                   
    \and
    K. Dohlen\inst{\ref{lam}}                  
    \and
    B.~Rouz\'e\inst{\ref{lam},\ref{centrale}}  
    \and
    J.-F.~Sauvage\inst{\ref{lam},\ref{onera}}  
    \and
    M.~El~Morsy\inst{\ref{lam}}                
    \and
    M.~Lopez\inst{\ref{lam}}                   
    \and 
    M.~N'Diaye\inst{\ref{oca}}                 
    \and
    A.~Caillat\inst{\ref{lam}}                 
    \and
    E.~Choquet\inst{\ref{lam}}                 
    \and
    G.~P.~P.~L.~Otten\inst{\ref{lam}}          
    \and 
    A.~Abbinanti\inst{\ref{lam}}               
    \and
    P.~Balard\inst{\ref{lam}}                  
    \and
    M.~Carbillet\inst{\ref{oca}}               
    \and
    P.~Blanchard\inst{\ref{lam}}               
    \and
    J.~Hulin\inst{\ref{lam}}                   
    \and
    \'E.~Robert\inst{\ref{lam}}                
}

\institute{
    Aix Marseille Univ, CNRS, CNES, LAM, Marseille, France\\
    \email{\href{mailto:raphael.pourcelot@oca.eu}{raphael.pourcelot@oca.eu}} \label{lam}
    \and
    Université Côte d’Azur, Observatoire de la Côte d’Azur, CNRS, Laboratoire Lagrange, France \label{oca}
    \and
    \'Ecole Centrale Marseille, France \label{centrale}
    \and
    ONERA, The French Aerospace Lab, BP72, 29 avenue de la Division Leclerc, 92322 Ch\^atillon Cedex, France \label{onera}
}

\date{\today}

\abstract
    {Imaging faint objects, such as exoplanets or disks, around nearby stars is extremely challenging because host star images are dominated by the telescope diffraction pattern. Using a coronagraph is an efficient solution for removing diffraction but requires an incoming wavefront with good quality to maximize starlight rejection. On the ground, the most advanced exoplanet imagers use extreme adaptive optics (ExAO) systems that are based on a deformable mirror (DM) with a large number of actuators to efficiently compensate for high-order aberrations and provide diffraction-limited images. While several exoplanet imagers with DMs using $\sim$1500 actuators are now routinely operating on large telescopes to observe gas giant planets, future systems may require a tenfold increase in the number of degrees of freedom to look for rocky planets. In this paper, we explore wavefront correction with a secondary adaptive optics system that controls a very large number of degrees of freedom that are not corrected by the primary ExAO system. Using Marseille Imaging Testbed for High Contrast (MITHiC), we implement a second stage of adaptive optics with ZELDA, a Zernike wavefront sensor, and a spatial light modulator (SLM) to compensate for the phase aberrations of the bench downstream residual aberrations from adaptive optics. We demonstrate that their correction up to 137\,cycles per pupil with nanometric accuracy is possible, provided there is a simple distortion calibration of the pupil and a moderate wavefront low-pass filtering. We also use ZELDA for a fast compensation of ExAO residuals, showing its promising implementation as a second-stage correction for the observation of rocky planets around nearby stars. Finally, we present images with a classical Lyot coronagraph on MITHiC and validate our ability to reach its theoretical performance with our calibration.
}


\maketitle

\section{Introduction}

Imaging extrasolar planets is one of the most demanding endeavors in astronomy today. Collecting photons from these planets will yield precious astrophysical information on their nature, leading to knowledge of their orbital and rotational parameters \citep{Galicher2014, Vigan2016, Chilcote2015, Maire2016, Zurlo2016}, clues on their location and formation history \citep{Crepp2012, Oberg2011, Piso2016}, and constraints on the chemistry at stake in their atmosphere \citep{Knutson2007, Phillips2020}. However, imaging an exoplanet, and in particular an Earth analog, is extremely challenging because of the flux ratio (also referred to as the contrast) of at least $10^{-8}$ with its host star at angular separations shorter than 50\,mas \citep{Traub2010}. Using a coronagraph on a large telescope is a compelling approach for attenuating the light of an observed star image and providing the sensitivity to collect sufficient photons from the planet at high angular resolution.

 For large telescopes on the ground, angular resolution does not scale with the telescope's primary mirror diameter as Earth's atmospheric turbulence blurs the observations. Adaptive optics (AO) systems have been developed with a wavefront sensor (WFS) and a deformable mirror (DM) to measure and compensate for the aberrations introduced by the atmosphere in real time, leading to images at nearly the theoretical resolution limit \citep[e.g.,][]{Beuzit1997,Herriot1998,Rousset2003,Roddier2004}. The accurate calibration of AO systems relies on the computation of the interaction matrix, a numerical tool relating the response of WFS subapertures to DM actuations.

The latest generation of high-contrast imaging instruments, such as the Very Large Telescope (VLT) Spectro-Polarimetric High-contrast Exoplanet REsearch (SPHERE) \citep{Beuzit2019}, Gemini Planet Image (GPI) \citep{2014PNAS..11112661M}, and the Subaru Coronagraphic Extreme Adaptive Optics (SCExAO) \citep{Jovanovic2015}, include novel extreme adaptive optics (ExAO). These systems are characterized by a wavefront correction with a high temporal frequency, up to 3.5\,kHz, and a DM with a large number of actuators, up to 40 across the telescope pupil diameter. While allowing for correction in a field of view up to $\sim$1\as$\times$1\as, ExAO enables the production of images with Strehl ratios up to 95\% in the near infrared \citep[e.g.,][]{Sauvage2016}. Such image quality translates into an increase in the amount of light gathered in the core of the point spread function (PSF) of the planet and, therefore, in the signal-to-noise ratio (S/N). In addition, the high image quality allows for efficient starlight suppression at large separations to observe faint structures, such as protoplanetary and debris disks, around an observed bright star.

These instruments have proven to be remarkably efficient, with contrasts down to $10^{-6}$ at 0.3\as angular separations, leading to several new young substellar mass companion discoveries \citep{Macintosh2015, 2015ApJ...799..182P, 2016ApJ...829L...4K, Chauvin2017, Keppler2018, Cheetham2018}, many newly imaged disks \citep[e.g.,][]{2018ApJ...863...44A, 2016A&A...595A.112G, 2016A&A...595A.113S, 2016A&A...595A.114D}, and studies of other astrophysical objects at unprecedented resolutions \citep[e.g.,][]{2015A&A...578A..77K,Schmid2017}. Extensive surveys such as the Gemini Planet Imager Exoplanet Survey (GPIES) with Gemini/GPI \citep{Nielsen2019} or the SpHere INfrared survey for Exoplanets (SHINE) with VLT/SPHERE \citep{Vigan2020} have brought new constraints on the population of young giant planets.

Their success has also shed light on the limiting factors that prevent observations at even deeper contrasts and smaller separations. One of the main limitations for coronagraphy are non-common path aberrations (NCPAs), which are optical distortions on the instrument science path that are unseen by the AO system on its WFS path. As they slowly vary with time \citep{Soummer2007}, these quasi-static aberrations do not average over time and cannot simply be calibrated as a bias. Over the past few years, several strategies have been proposed to calibrate these residual errors (see, for example, the review by \citealt{Jovanovic2018}). 

Among these methods, the Zernike wavefront sensor \citep[ZWFS;][]{1934MNRAS..94..377Z, 2003SPIE.5169..309B,Dohlen2004,2011SPIE.8126E..0FW} is one of the leading approaches for NCPA correction in exoplanet imagers. Based on this approach, the Zernike sensor for Extremely Low-level Differential Aberration \cite[ZELDA;][]{N'Diaye2013} was developed, installed, and tested on VLT/SPHERE, showing promise as a calibration tool \citep{N'Diaye2016,Vigan2019}. This device allowed for diagnosing the low-wind effect, which has been crippling early SPHERE observations \citep{Sauvage2015}. It has also been explored in simulations for alternative applications, such as fine telescope segment co-phasing \citep{Janin-Potiron2017}. The ZWFS is set to be implemented on the Nancy Grace Roman Space Telescope, a future NASA flagship mission \citep{Shi2016} with high-contrast capabilities for exoplanet observations, as well as in future ground-based extremely large telescope (ELT) instruments \citep[e.g.,][]{Carlotti2018}.

When looking at the instrumentation requirements for the ELTs currently being built, there is a clear trend for larger numbers of actuators. For example, the M4 mirror in  the European Southern Observatory (ESO) ELT\footnote{\url{https://www.eso.org/sci/facilities/eelt/}} will have approximately 80 actuators across the pupil diameter \citep{Vernet2012}. This DM will provide an actuator pitch of $\sim$50\,cm and run at a maximum rate of 1\,kHz, which will be sufficient to provide images with a $\sim$70\% Strehl ratio in the $K$ band for the first light instruments \citep{Thatte2016,Davies2016} in single-conjugated AO mode. While this will enable detection capabilities for studying warm or massive gaseous planets \citep{Carlotti2018,Houlle2020inprep}, it will not be sufficient to provide ExAO performance for observing cold and light rocky planets \citep{Kasper2010}. For such a challenging task, an actuator pitch of $\sim$20\,cm and an AO loop running at 2-3\,kHz is required \citep{Kasper2011,Kasper2013}, resulting in nearly 200 actuators across the pupil diameter. Similarly, and although it is composed of several independent primary mirrors and associated deformable secondary mirrors, the Giant Magellan Telescope \citep{Johns2012} will provide a pitch of $\sim$20\,cm at first light with close to 120 actuators across the pupil diameter. The number of actuators also tends to increase for space applications. For example, \citet{Ruane2020} performs phase conjugation with two DMs from Boston Micromachines, yielding 50 actuators across the diameter of the active area for each device. It is therefore crucial to investigate the wavefront correction possibilities that are offered with a very large number of degrees of freedom.

To handle this new range of high-order spatial frequencies, we explore the use of a second-stage AO system, including a ZWFS and a spatial light modulator (SLM) implemented downstream from the first ExAO system. To make the most of the ZWFS formalism developed by \citet{N'Diaye2013}, we propose the use of a phase conjugation technique with a calibration procedure that uses probes that are performed in a two-frame acquisition sequence. We thus do not compute any standard interaction matrix: Each iteration implies solving a second-order polynomial equation per phase measurement bin.

In Sect.~\ref{sec:mithic}, we present the Marseille Imaging Testbed for High Contrast (MITHiC) with its 2019 upgrade and a ZELDA prototype to test the measurement and correction on a very large number of degrees of freedom. We also describe the closed-loop tests using ZELDA as a WFS. In Sect.~\ref{sec:phaseconj}, the experimental procedure is detailed with the ZELDA formalism, the Fourier filtering process, and the two-frame calibration of pupil distortion. In Sect.~\ref{sec:results}, we present the test results for static phase error correction and for an on-bench simulation of real-time SPHERE residual turbulence. Under certain assumptions, we show that it is possible to measure and compensate for aberrations down to a few nanometers in standard deviation in the optical path difference (OPD) with various input phase errors.

\section{The MITHiC test bed}
\label{sec:mithic}

\begin{figure*}
    \centering
    \includegraphics[width=0.9\textwidth]{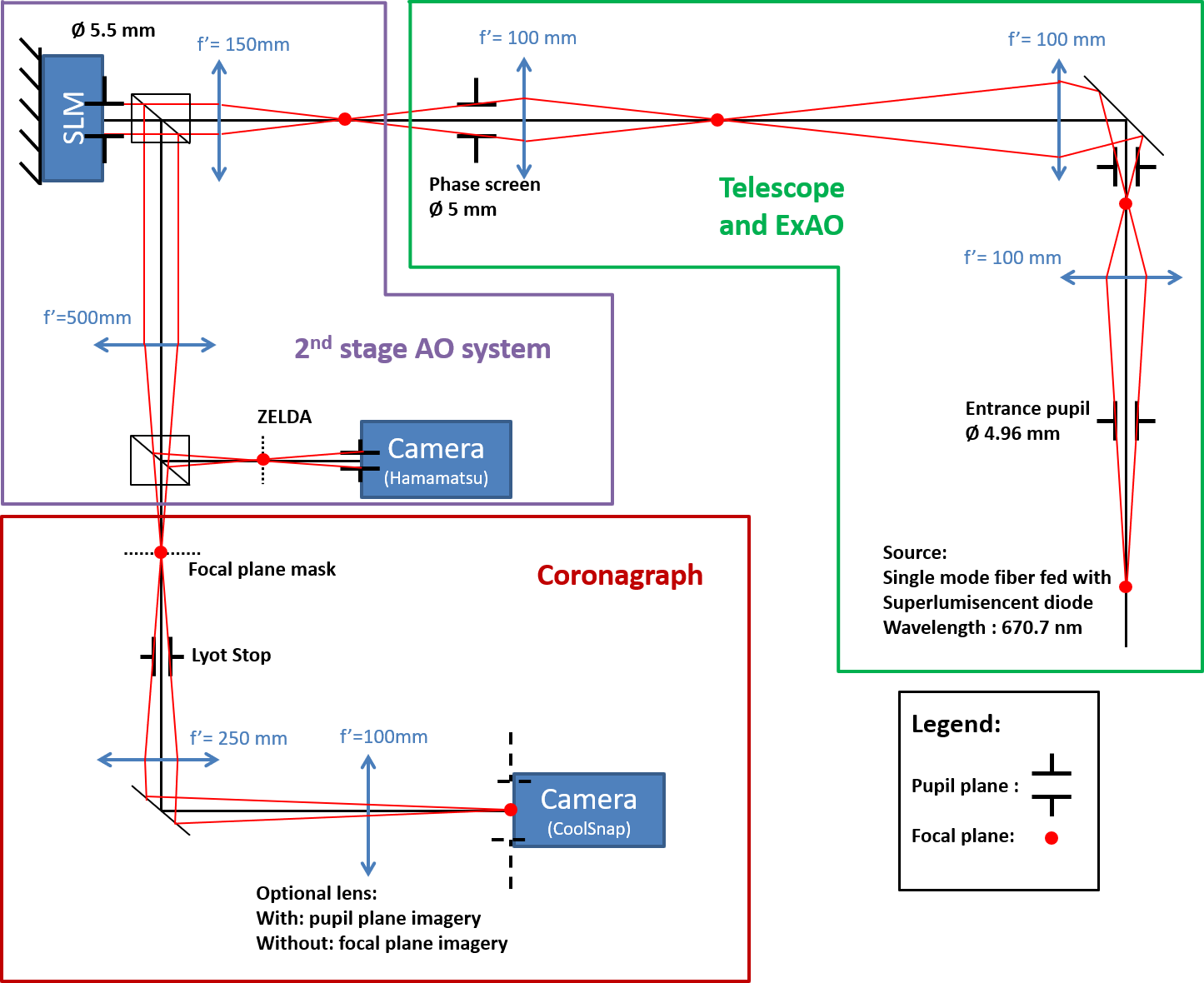}
    \caption{Schematic view of the MITHiC test bed. Focal planes are marked with red dots, and pupil planes are marked with black apertures with the corresponding pupil diameters. Beam splitter cubes are drawn in black. The telescope, WFS, and coronagraph parts are detailed in Sects.~\ref{sec:telescope}, \ref{sec:zelda}, and~\ref{sec:corono}, respectively. In addition, a control computer (not represented) is used to control the bench; some details are provided in Sect.~\ref{sec:control}.}
    \label{fig:mithic}
\end{figure*}

MITHiC was initiated in 2010 to develop and test technologies for exoplanet high-contrast imaging \citep{N'Diaye2012,N'Diaye2014}. It was used as a test bed for wavefront sensing techniques such as ZELDA \citep{N'Diaye2013} or COronagraphic Focal-plane wave-Front Estimation for Exoplanet detection (COFFEE) \citep{2013OExpr..2131751P}, coronagraphic concepts such as the apodized pupil Roddier \& Roddier or the dual-zone phase mask coronagraphs \citep{Soummer2003a,Soummer2003b,N'Diaye2010,N'Diaye2012}, and dark hole algorithms \citep{herscovicischiller:tel-02015796,leboulleux:tel-02167863}, as well as other science, such as detector characterization \citep{2019arXiv191000374G}. Figure~\ref{fig:mithic} presents the current optical setup of the bench. It is divided into three major parts, which are briefly described here and in much greater detail in Appendix~\ref{sec:mithic_detailed}.

The telescope part goes from the light source to a dedicated phase screen and simulates a telescope observing an unresolved star. The light source is a fiber-coupled superluminescent diode, which has a central wavelength $\lambda = 670.7$\,nm and a spectral bandwidth of 10\,nm. The source is linearly polarized to match the SLM requirements precisely. A phase screen manufactured by SILIOS Technologies is then used to introduce various types of aberration patterns that are typically encountered in large optical telescopes. It includes both static patterns and two continuous rings of simulated residual turbulence for VLT/SPHERE and the ELT High Angular Resolution - Monolithic - Optical and Near-infrared - Integral field spectrograph (HARMONI) \citep{Vigan2016SPIE}.

The following part is the second-stage AO system that is based on an SLM for the phase correction and a ZELDA WFS. The pupil of the system is reimaged on an SLM that mimics a DM to apply phase corrections. Our implementation allows for a particularly large number of subapertures on the WFS and a very high density of correction elements on the SLM, with, respectively, 418 measurement points and 274 pixels across the pupil diameter, $D$. As a downside, using an SLM as a wavefront corrector imposes the use of linearly polarized monochromatic light. The ZELDA WFS is implemented using a phase mask with a depth of 350\,nm, corresponding to nearly $\lambda/4$ in optical depth at 670.7\,nm, and a diameter of 64\,\mic, corresponding to a relative size of 1\lsd \citep{N'Diaye2013}. It is mounted on a three-axis mount that enables its positioning with 1\,\mic accuracy. The pupil is then reimaged with a diameter of 418 pixels on the detector of ZELDA, which provides more than 130,000 single-resolution elements for wavefront sensing. The acquisition procedure and the associated formalism for wavefront measurements are detailed in Sect.~\ref{sec:phaseconj}. The WFS can be operated for single wavefront measurements, but it is best used in a closed loop with the SLM, as presented in the following sections. Without any correction from the SLM, the total amount of aberrations is approximately equal to 35.5\,nm\,root mean square (RMS) on the wavefront and is mainly due to the spherical aberration from cumulative wavefront errors from the lenses. Using the phase error compensation described in Sect.~\ref{sec:phaseconj}, the level of residual phase aberrations can ultimately be reduced down to $\sim$2\,nm\,RMS, which is close to the theoretical quantization limit of the SLM at 0.76\,nm\,RMS (see details in Sect.~\ref{sec:mithic_detailed}).

Finally, the last part of the bench corresponds to the coronagraph and associated science imaging channel. The test bed currently implements a classical Lyot coronagraph (CLC) based on an opaque focal plane mask (FPM) and an associated Lyot stop. In the final focal plane, the science imaging camera provides a fast readout with low noise to obtain high-contrast images. The camera can be used to image the coronagraphic signal either in the focal plane or in the pupil plane with an optional lens.

\section{Principle of wavefront correction in a closed loop } 
\label{sec:phaseconj}

We used the bench to perform closed-loop corrections of various types of phase errors. The phase measurements were performed with ZELDA, whose principle is recalled in Sect.~\ref{sec:zwfs}. As this sensor naturally provides phase error maps, we worked with an interaction-matrix-free correction by applying the opposite measured phase map to the SLM. The algorithm is described in Sect.~\ref{sec:procedure}. To overcome distortion in pupil conjugation, which is automatically dealt with in a classical system relying on an interaction matrix, we performed an additional calibration step, detailed in Sect.~\ref{sec:calibration}. We finally performed an additional low-pass spatial filtering to avoid aliasing propagation and calibration errors (see Sect.~\ref{sec:filtering}). 

\subsection{Phase computation with ZELDA} 
\label{sec:zwfs}

ZELDA uses a focal plane phase mask on the point source image to produce interferences between the light going through and surrounding the mask in a downstream pupil plane. This results in pupil intensity variations, $I_c$, that are related to the phase aberrations, $\varphi$, upstream from the phase mask. In the regime of small aberrations ($\varphi\ll$1\,rad) and assuming a second-order Taylor expansion for $\varphi$, the intensity--phase relation from \citet{N'Diaye2013} yields
\begin{equation} \label{eq:zernike_formula}
    I_c = P^2 + 2b^2(1-\cos{\theta}) + 2Pb\left[\varphi \sin{\theta} - (1-\varphi^2/2)(1-\cos{\theta})\right],
\end{equation}
where $\theta$ is the phase shift introduced by the mask, $P$ denotes the amplitude profile in the entrance pupil (a top-hat function in our case), and $b$ represents the real amplitude profile, which only depends on the size of the ZELDA mask and can be calculated once and for all. Retrieving the phase map is obtained by solving a second-degree polynomial equation for each pixel in the detector pupil plane. The OPD map is simply derived by inverting the relation $\varphi=2\pi OPD/\lambda$. This formalism is cost-efficient in terms of computation.

\subsection{Aberration compensation on MITHiC}
\label{sec:procedure}

On MITHiC, the pupil is sampled with 145,000 subapertures in the pupil on the WFS camera and more than 60,000 DM-like actuators on the SLM. With such large numbers of degrees of freedom, we implemented the direct correction of the aberrations without the requirement for a full-fledged interaction matrix; it would be extremely time-consuming to compute on MITHiC, which is not optimized for high-speed operations.

The first step is to acquire an initial OPD map, which requires two acquisitions on the WFS camera: one with the ZELDA mask and one without it for flux normalization, as described in \citet{N'Diaye2013}. Both images are dark-corrected and re-centered. The OPD map is then computed using Eq.~(\ref{eq:zernike_formula}). In practice, we use the \texttt{pyZELDA}\footnote{\url{https://github.com/avigan/pyZELDA}} package for the phase reconstruction \citep{2018ascl.soft06003V}. With 418 pixels across the pupil diameter, the wavefront measurement with ZELDA allows us to retrieve extremely fine phase information with spatial frequencies up to 209\,cycles per pupil (c/p).

Since the pupil is sampled with 1.5 times more pixels by the ZELDA camera than the SLM, we first tried to directly apply the negative of a resampled phase map onto the SLM to compensate for the measured aberrations. However, as we will see in Sect.~\ref{sec:results}, this leads to a poor wavefront correction in a closed loop due to the geometrical distortion of the pupil. We therefore perform an additional calibration, as well as a spatial filtering of the computed OPD. 
Once these calibrations are applied, the processed OPD map is downscaled using spline interpolation to get a pupil diameter of 274 pixels to match the pupil size on the SLM, converted into phase, and displayed on the SLM.

For phase errors larger than 40\,nm\,RMS, the ZELDA wavefront reconstruction is outside the sensor linearity range and the error is underestimated \citep{N'Diaye2016}. Therefore, a single correction step will, in general, not be sufficient to fully compensate for the phase errors of the bench. Repeating the previous procedure in a closed-loop fashion allows for convergence in a small number of iterations.

\subsection{Compensation for the geometrical distortions}
\label{sec:calibration}

\begin{figure*}
    \centering
    \includegraphics[width=\linewidth]{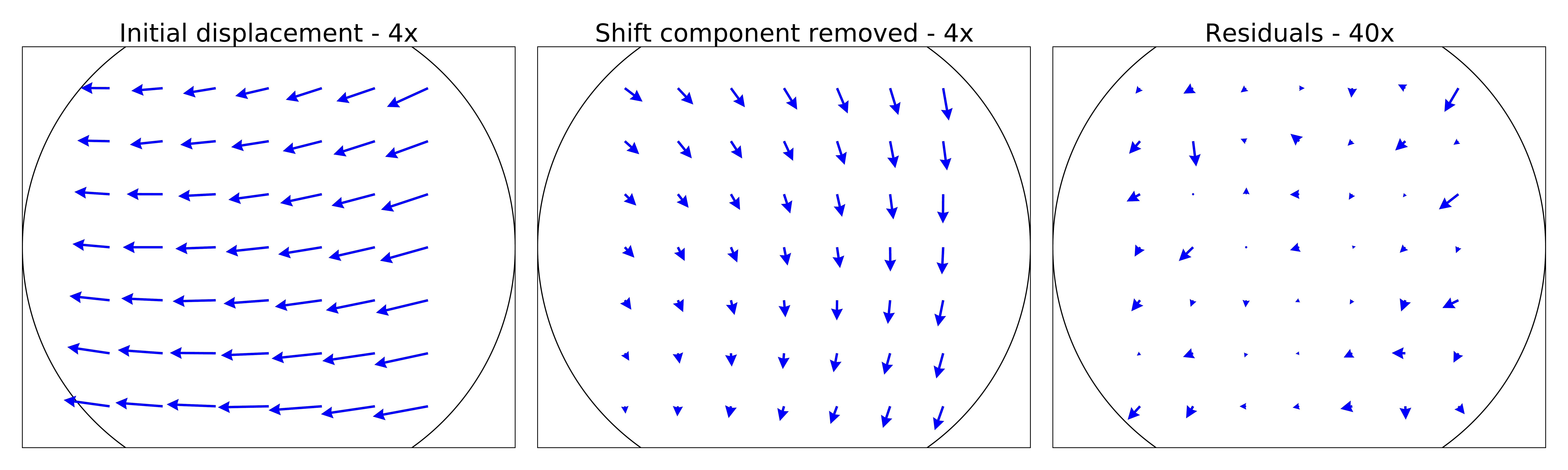}
    \caption{Geometrical distortion of the pupil between the SLM and the ZELDA WFS. The left plot shows the full pixel displacement between the introduced Gaussian spots and their measured positions, with a scale multiplied by four.\ The center plot shows the displacement after subtracting the $\textbf{C}$ component, again with a scale multiplied by four. The right plot shows the final residual displacement after correction of the measured positions, this time with a scale multiplied by 40. The black circles represent the pupil footprint.}
    \label{fig:dist_plots}
\end{figure*}

In the absence of a classical interaction matrix, we calibrated the system to compensate for geometric pupil distortion between the WFS camera and the SLM. The calibration step was performed by introducing a grid of $N\times N$ Gaussian phase spots on the SLM, for which we know the exact theoretical position. The input coordinates, $x_i, y_i$, of each spot, $i$, can be represented by a column vector, $\vec{V_i}$. In the OPD map measured with ZELDA, we then retrieve the output position vector of the spots, $\vec{V'_i}$. If we consider an affine distortion, there exists a $2\times2$ matrix, $\vec{M}$, and a $2\times1$ column vector, $\vec{C}$, such that, for all spots $i$, the $\vec{V_i}$ coordinates satisfy:
\begin{equation}
    \vec{V_i} = \vec{M}\,\vec{V'_i} + \vec{C}\,.    
\end{equation}

We fit this transformation using a least-square algorithm to determine the  $M$ and $C$ coefficients to solve the following optimization problem:
\begin{equation}
\begin{aligned}
    & \underset{\vec{M}, \vec{C}}{\text{min}} 
    & \sum_{i=1}^{N}\left(\vec{V_i} - \vec{M}\,\vec{V'_i} + \vec{C}\right)^2\,.
\end{aligned}
\end{equation}

Finally, using the estimated $\hat{\vec{M}}$ and $\hat{\vec{C}}$ from the fit, we applied the inverse transform to each spot of our OPD maps to derive $\vec{V'_i}$:
\begin{equation}
    \vec{V'_i} = \hat{\vec{M}}^{-1}\,\left(\vec{V_i} - \hat{\vec{C}}\right)\,.
\end{equation}

Figure~\ref{fig:dist_plots} illustrates this process with a $7\times7$ grid. This square grid does not probe all of the surface of the pupil disk but only an inscribed square centered in the pupil. Nevertheless, the 49 introduced dots already ensure the convergence of the determination of the six parameters of $\textbf{M}$ and $\textbf{C}$. A more complete sampling would have to be tested even if it might not yield increased precision. The computing cost of this calibration sums up to: two frame acquisitions, 7$\times$7 Gaussian fits on 32\,$\times$32\,pixel subarrays, and a linear regression to fit a six-parameter function on 49 points, which is performed within a few tens of seconds. Despite the fact that the bench is not stabilized in temperature or humidity, we found that performing the calibration once per day is sufficient for our usage. Figure~\ref{fig:dist_plots} (left) shows a clear difference between the introduced calibration spots (arrow ends) and the measured calibration spots  (arrow tips). The error shows not only a shift to the left, which can be seen as a misalignment, but also an additional, more complex distortion pattern. When removing the pure shift component, as in Fig.~\ref{fig:dist_plots} (center), the maximum distance between the introduced and measured spot positions is 6.96 pixels on the ZELDA camera, with an average of 4.27 pixels. In terms of pupil percentage, this average displacement writes as $0.01D$, which is a sub-actuator displacement for a VLT/SPHERE-like DM. The affine approximation to compensate for this error provides accurate results as the position of the spots after correction closely matches the expected position. The maximum distance between the introduced and corrected spots is shorter than 0.6 pixels, which corresponds to less than $0.002D$. The final displacements are presented in Fig.~\ref{fig:dist_plots} (right). The average distance between an introduced point and a corrected point is 0.15 pixels, which shows an efficient correction. Different grid sizes were tested, but $N$ larger than 7 do not provide any quantitative improvements of the distortion correction; as such, we use this value from here on out. This calibration method can be easily implemented on any instrument already hosting a Zernike WFS, such as VLT/SPHERE \citep{Beuzit2019} and the Roman Space Telescope (RST) Coronagraphic Instrument (CGI) \citep{Kasdin2020}, or on test beds, such as the High Contrast Imaging Testbed (HCIT) \citep{Ruane2020}. This calibration is, in fact, applicable to any wavefront sensor that provides measurements in the pupil plane, such as the pyramid WFS \citep{Ragazzoni1996}, or more generally with most of the class of Fourier-based WFSs \citep{Fauvarque2016}.

\subsection{Low-pass filtering}
\label{sec:filtering}

\begin{figure}
    \centering
    \includegraphics[width=\columnwidth]{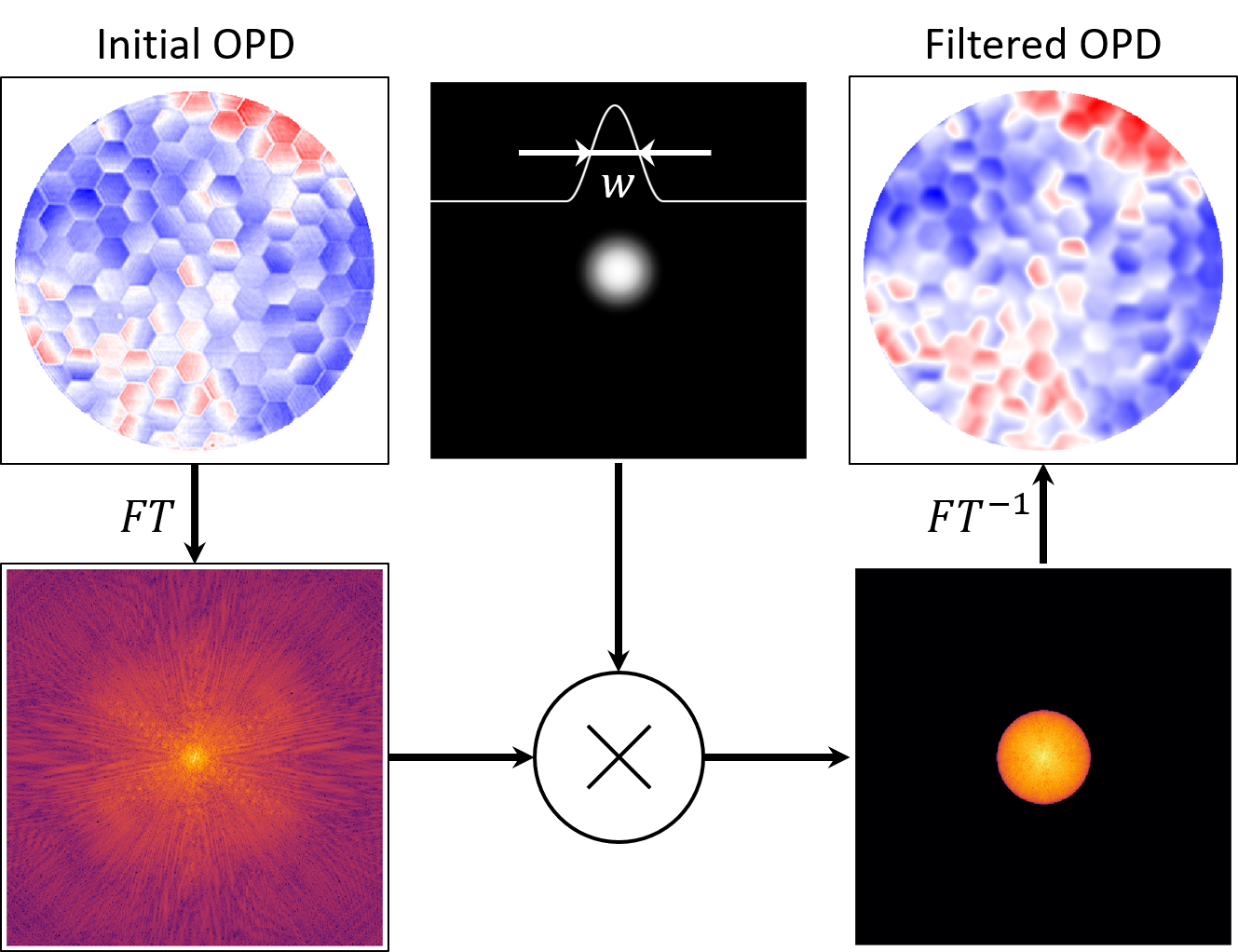}
    \caption{Detailed steps of Fourier filtering. A Fourier transform (FT) is first applied to the initial OPD. The array in Fourier space is multiplied by the Hann window of total width $2w\,\lambda/D$. Finally, an inverse FT, denoted FT\textsuperscript{-1}, is applied to retrieve the filtered OPD.}
    \label{fig:fourier_filtering}
\end{figure}

Aliasing effects will appear when a wavefront measurement at very high spatial frequency is applied on a much lower-order DM (or equivalent). This was previously highlighted for NCPA correction with ZELDA in VLT/SPHERE, where the OPD maps had to be either low-pass filtered in the Fourier space \citep{N'Diaye2016} or alternatively projected onto the controllable modes of the ExAO system \citep{Vigan2019} before being applied on the DM. In the present case, we implemented an optional low-pass filtering procedure by multiplying the Fourier transform of the OPD map by a Hann window of a specific width, as illustrated in Fig.~\ref{fig:fourier_filtering}. We adapted the window width, $w$, to match the desired cutoff frequency. As a definition of the filter value, we set $w$ to be the spatial frequency value at which the Hann window reaches zero. Therefore, a low-pass filter at $w$\,c/p yields a Hann window diameter of $2w\lambda/D$.

\begin{figure}
    \centering
    \includegraphics[width=\columnwidth]{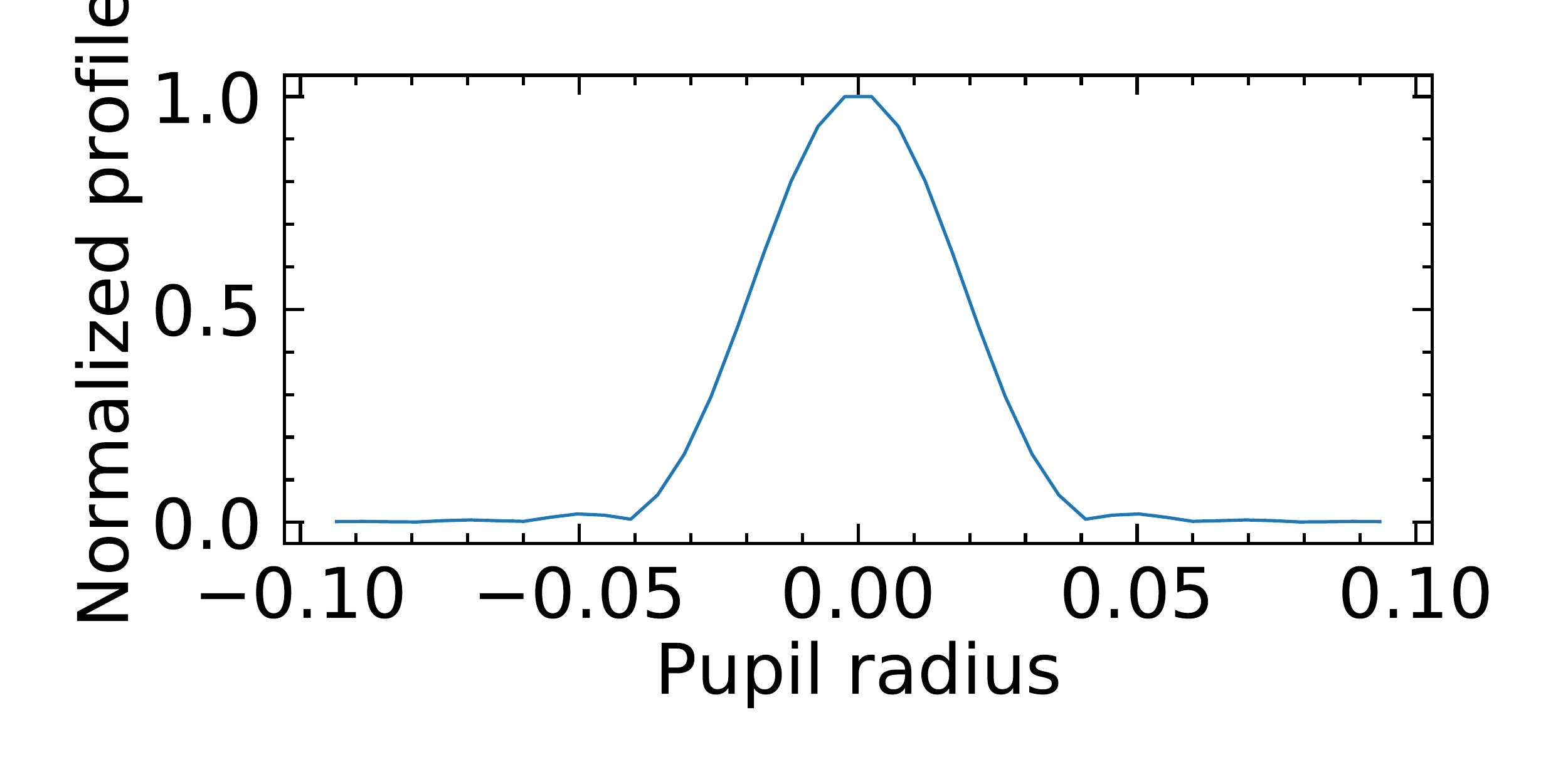}
    \caption{Illustration of the radial profile of an influence function corresponding to a Hann filter with $w=$52\,c/p, the equivalent of a DM with 104 actuators across.}
    \label{fig:influence_funct}
\end{figure}

Using this method of filtering is analogous to choosing the number of actuators and the influence function of the equivalent DM. With a filter cutoff frequency at $w$\,c/p, we modeled a DM with $2w$ actuators. With a Hann window, the full width at half maximum of the corresponding actuators' influence function is $D/2w$. Figure~\ref{fig:influence_funct} highlights the radial profile of the corresponding influence function, computed by a Fourier transform of the filtering window. The influence function is Gaussian-like and is therefore a good approximation of the influence function of classical DMs  based on either piezo-stack actuators or micro-electro-mechanical systems (MEMS). We could have chosen different filter functions for more realistic simulations of specific DM architectures. 

\section{Results} 
\label{sec:results}

In this section we present our results in different configurations. We demonstrate first a static configuration that corresponds to the compensation for various static phase maps, and second, a dynamic configuration where we introduce residual atmospheric aberrations representative of the on-sky performance of the SPHERE ExAO (SAXO) system. In this configuration we tried to correct for the ExAO residual aberrations in pseudo-real time by using the tools developed in the static case. Finally, we looked at the improvement in MITHiC coronagraphic images following a fine residual aberration compensation. 

For all the results hereafter, the tip and tilt aberrations have been removed. While ZELDA is highly sensitive to these modes, in most AO systems they are managed by a dedicated tip and tilt sensor and/or a dedicated mirror that runs at a very high frequency. Furthermore, MITHiC is affected by a small amount of thermal turbulence inside its enclosure, which induces tip and tilt variations of typically 2--3\,nm\,RMS in the 5--10\,Hz frequency range. These variations appear at a higher frequency than our correction loop (currently $\sim$1\,Hz), so we do not address them directly in the present work.

\subsection{Static aberrations} 
\label{sec:static}

\begin{figure*}
    \centering
    \includegraphics[width=\textwidth]{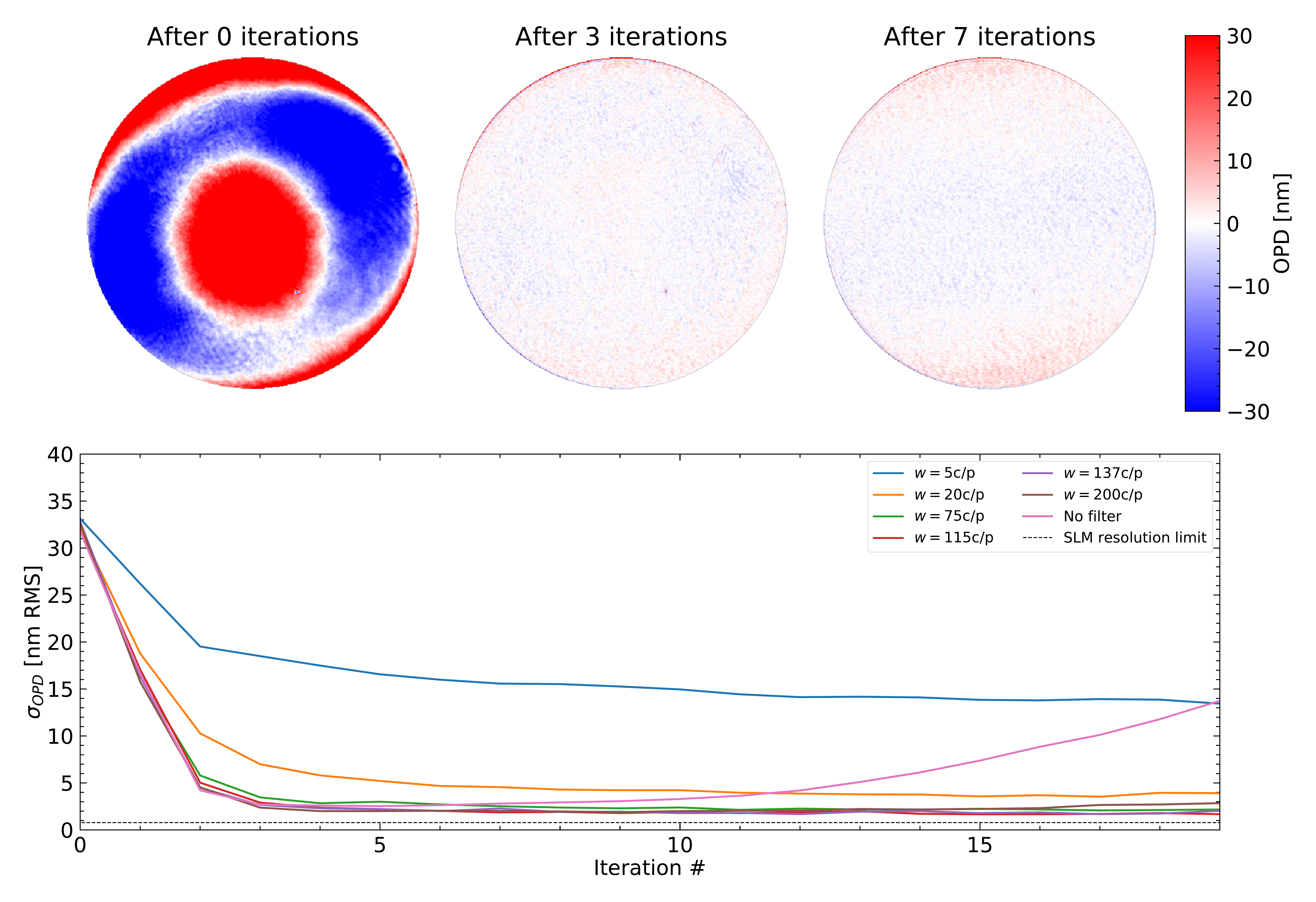}
    \caption{Bench residual aberration compensation in a closed loop with the pupil distortion correction and a filter of 115\,c/p. The top row shows wavefront measurements obtained with ZELDA at iterations 0 (initial aberration), 3, and 7. The bottom plot shows the OPD standard deviation in the pupil as a function of iteration number and for various values of filtering. A filter value of 20\,c/p (orange) simulates a VLT/SPHERE-like DM cutoff filter. For each measurement, the global tip and tilt are subtracted.}
    \label{fig:ncpa_cal}
\end{figure*}

We addressed the correction of the quasi-static aberrations on MITHiC, which are illustrated in Fig.~\ref{fig:ncpa_cal} (top left). The standard deviation in the OPD inside the pupil, $\sigma_{OPD}$, is typically 35\,nm\,RMS at the beginning of the correction step. With the application of the closed-loop correction procedure, including the geometrical distortion correction and for various values of $w$, we obtained the results shown in Fig.~\ref{fig:ncpa_cal}. The top plot shows the OPD maps from ZELDA at three representative iterations of the procedure with $w=$115\,c/p. The bottom plot displays the evolution of $\sigma_{OPD}$ at different iterations.

The general behavior is identical for most values of $w$, with a steady decrease in $\sigma_{OPD}$ during the first two or three iterations followed by a stabilization during the following iterations. This is due to the limited linearity range of ZELDA, which is reduced compared to other classical WFSs, such as the Shack-Hartmann WFS. As highlighted in \cite{N'Diaye2013}, the linear behavior of ZELDA is typically limited to $\pm0.03\lambda$ ($\pm20$\,nm here), and larger absolute values of phase error will always be underestimated by the second-order reconstruction. As long as there is no phase wrapping, the estimation will always favor a decrease in the wavefront error, which means that a correction in a closed loop will eventually converge. The same behavior is used for co-phasing techniques proposed in multi-wavelength approaches with the ZWFS \citep{Vigan2011,Cheffot2020}. Ultimately, the usable capture range of ZELDA in a closed loop depends on the phase shift imposed by the mask and ranges between $-0.14\,\lambda$ and $+0.36\,\lambda$ for the MITHiC prototype \citep{N'Diaye2016}.

For $w$ between 75 and 137\,c/p, $\sigma_{OPD}$ stabilizes around $\sim$1.9\,nm\,RMS, which is above the theoretical limit of 0.76\,nm\,RMS. Most of the residual aberrations can be attributed to low-order aberrations from the internal turbulence on MITHiC, vibrations of the bench, and dust particles on optical surfaces that create small errors in the ZELDA reconstruction. For $w=$20\,c/p, the convergence is slower than for higher filtering values and it does not quite reach this low value of 1.9\,nm\,RMS, which indicates that there is a non-negligible amount of aberrations with spatial frequencies between 20 and 75\,c/p. The $w=$5\,c/p case is even more extreme, with an even slower decrease in $\sigma_{OPD}$ and a convergence toward a larger value of $\sim$14\,nm\,RMS after 15 iterations. 

The $w=$200\,c/p and the no-filter curves show the same steady decrease during the first iterations, showing that the filter can be optional for up to five iterations, for example to record a flat position to perform other science later. However, the loop is no longer stable and diverges after about seven iterations when no filter is applied and after 15-20 iterations for $w=$200\,c/p. We note that, even though the $w=$200 c/p filter maintains all the spatial frequencies measured by ZELDA, the attenuation of the highest frequencies by the apodization of the filtering window allows for a significant stabilization of the loop. We attribute this effect to a misestimation or mis-correction of the highest spatial frequencies -- which originate from pupil edge effects aliasing since the correction maximum frequency is at 137\,c/p -- as well as higher-order distortion residuals since we only calibrate for linear transformations.
 
\begin{figure*}
    \centering
    \includegraphics[width=\textwidth]{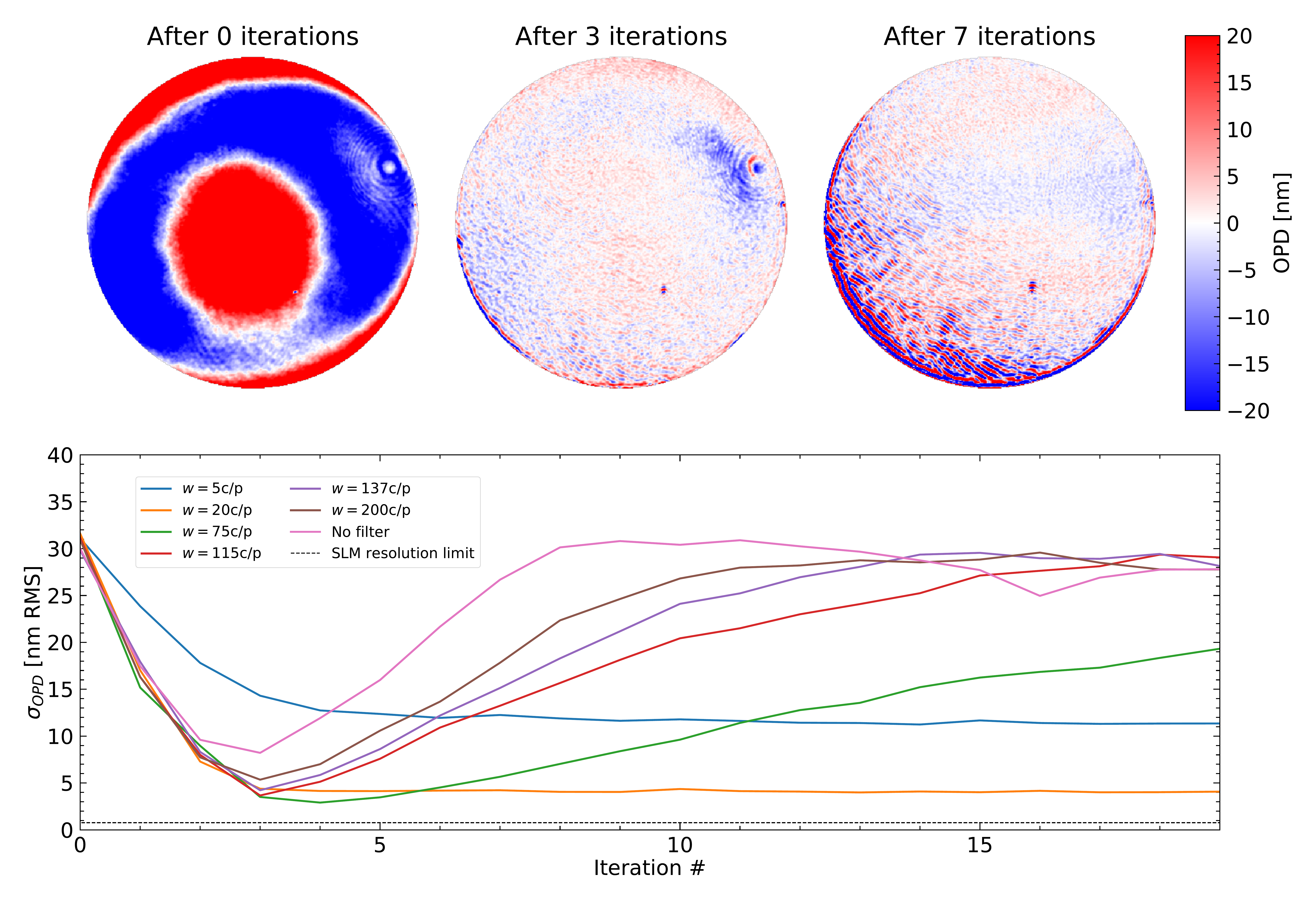}
    \caption{Bench residual aberration compensation in a closed loop without any correction of the pupil distortion and with a filter of 115\,c/p. The top row shows wavefront measurements with ZELDA at iterations 0 (initial aberration), 3, and 7. The bottom plot shows the OPD standard deviation in the pupil as a function of iteration number and for various values of filtering. The filter value of 20\,c/p (orange) simulates a VLT/SPHERE-like DM cutoff filter. For each measurement, the global tip and tilt are subtracted.}
    \label{fig:ncpa_nocal}
\end{figure*}

To emphasize the need for the pupil distortion calibration detailed in Sect.~\ref{sec:calibration}, we present the convergence results without the distortion correction in Fig.~\ref{fig:ncpa_nocal}. In these tests, the measured OPD map is simply filtered, scaled down, and applied to the SLM. For the two smallest $w$ (5 and 20\,c/p), the results are strictly identical to the previous test. For larger $w$, however, the behavior is very different: After reaching a minimum around 5\,nm\,RMS, the standard deviation quickly diverges and converges to values larger than 20\,nm\,RMS after ten iterations. 

The divergence is caused by the apparition of high-spatial-frequency wrinkles that slowly degrade the wavefront quality. The issue remains insignificant for $w=$20\,c/p, corresponding to a VLT/SPHERE-like case, but becomes critical for $w$ larger than 75\,c/p. If the computation of an interaction matrix can be performed to get rid of these ripples, new adaptive ELTs might struggle to construct them due to the absence of an appropriate calibration source \citep{Heritier2018}. Our calibration therefore shows another way of sensing the distortion modes and improving the correction with on-sky measurements with a simple implementation as long as the phase errors are not too large. In systems with large phase errors, such as a system without a first step of AO correction, further strategies would have to be explored. For example, to measure the distortion, the position of the introduced Gaussian spots could be estimated through the turbulence by averaging turbulence with long exposures. Further studies would have to be performed to assess the precision of such measurements.

\begin{figure*}
    \centering
    \includegraphics[width=\textwidth]{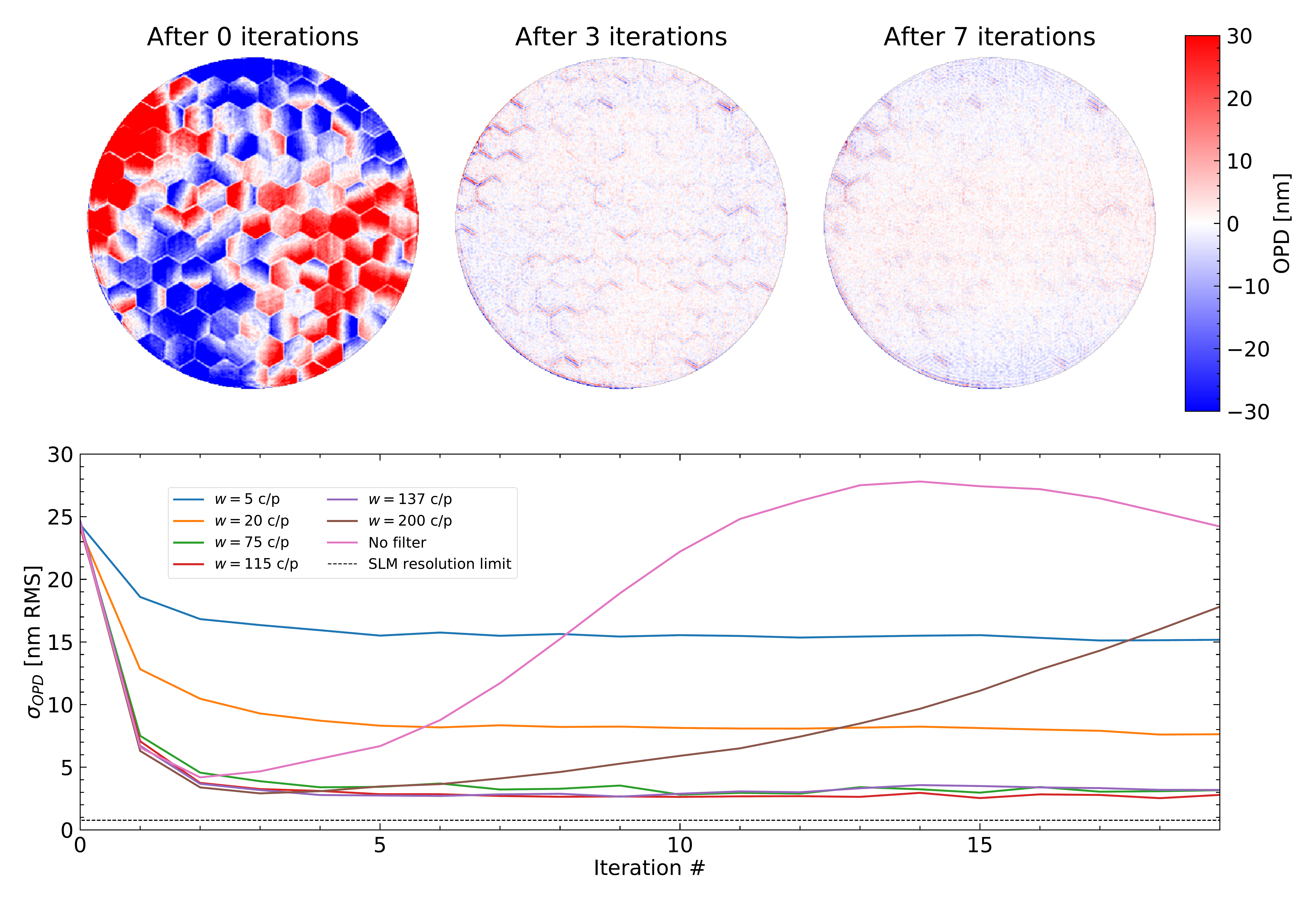}
    \caption{Compensation in a closed loop of a pattern simulating co-phasing errors of a segmented mirror using a low-pass filter value of 115\,c/p. The pattern is one of the static configurations of the MITHiC phase screen (see Sect.~\ref{sec:telescope}). The top row shows wavefront measurements obtained with ZELDA at iterations 0, 2, and 6. The bottom plot shows the OPD standard deviation in the pupil as a function of iteration number.}
    \label{fig:coph}
\end{figure*}

Based on the successful implementation of our wavefront correction procedure, we highlight one emblematic case of quasi-static aberrations where a very large number of degrees of freedom for wavefront correction would be particularly interesting for future applications: the compensation for residual co-phasing errors for telescopes with segmented primary mirrors. To test ZELDA in this practical case, we selected a pattern that combines typical NCPAs and a segmented pupil with random tip, tilt, and piston on individual segments on the MITHiC SILIOS phase screen. The results of the correction procedure with $w=$115\,c/p are illustrated in Fig.~\ref{fig:coph}. The 3\,nm\,RMS residual wavefront error after a few iterations shows the ability of ZELDA  to measure fine differential piston errors. The high 115\,c/p cutoff value for $w$ ensures a good correction for the segment discontinuities.

The use of ZELDA for fine segment co-phasing was proposed by \citet{2017A&A...603A..23J}; in their approach, they measure the residual piston, tip, and tilt of individuals segments and then apply the correction directly on the segments using their control actuators. In our case, as we benefit from a large number of subapertures and actuators, we go one step further and correct not only for the piston, tip, and tilt, but also for higher-order phase errors such as low-order aberrations on individual segments. On a pupil with 30 segments across the diameter, such as the ELT, 120 actuators across the pupil diameter will yield roughly 4x4 actuators per segment, which would allow for subsegment aberration correction. The curve in Fig.~\ref{fig:coph} that is the closest to this case is the orange curve, simulating a case with 40 actuators for 11 segments. This sort of possibility could prove extremely valuable for exoplanet imagers on future ELTs.

\subsection{Temporal error analysis with residual ExAO aberrations }
\label{sec:fast_zelda}

\begin{figure*}
    \centering
    \includegraphics[width=\textwidth]{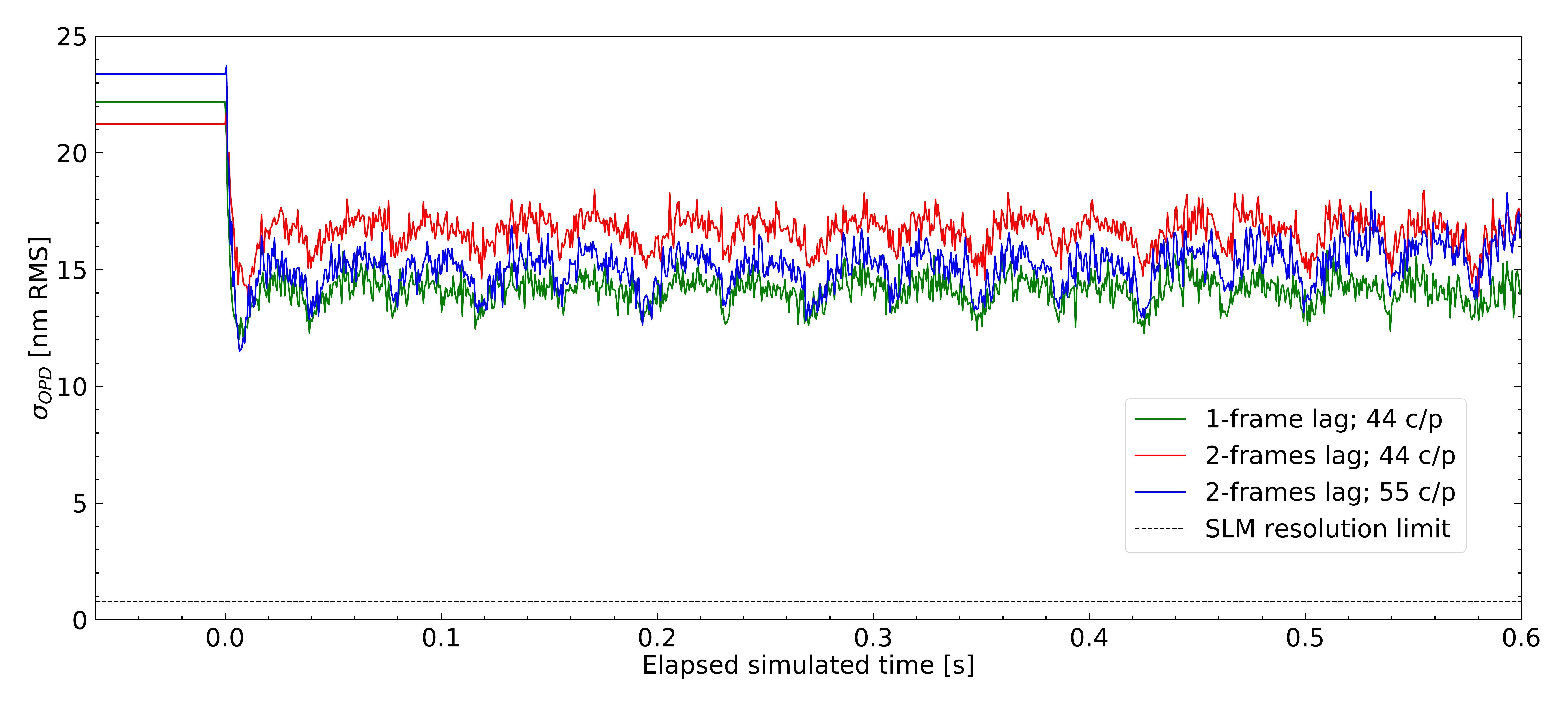}
    \caption{Evolution of the standard deviation in the OPD as a function of iteration when correcting VLT/SPHERE residual atmospheric turbulence with a wind speed of 10\,m/s and an iteration lag of 0.67\,ms for 0.6\,s (1000 iterations). The green curve shows the case with a filter at 44\,c/p and a one-frame lag, the red curve with a filter at 44\,c/p and a two-frame lag, and the blue curve with a filter at 55\,c/p and a two-frame lag.  The correction loop starts at time 0.}
    \label{fig:wind_speed10}
\end{figure*}

When using ZELDA, it is also worth considering the compensation for turbulence residuals in real time, for example downstream an ExAO system that provides first-stage correction. Similarly to the co-phasing case, this approach could potentially provide a second-stage WFS to increase the ExAO correction level in extremely demanding cases.

The phase screen installed on MITHiC provides an external ring of engraved turbulence that simulates residuals from the VLT/SAXO system \citep{Fusco2006} in median seeing conditions at the Paranal Observatory. This ring can be rotated at a controlled speed to emulate the wind speed. To overcome the bench limitation in terms of temporal frequency ($\sim$1\,Hz), we discretized the steps to simulate a VLT/SPHERE residual turbulence for an ExAO system, which, in practice, means that the screen rotates very slowly and in discrete steps that are synchronized with the ZELDA measurements. The delay associated with one-frame processing is therefore 0.67\,ms.

To explore several simple, yet realistic, AO systems, we included a delay, $d$, expressed in numbers of frames. The turbulence starting point on the phase screen was identical for each experiment. At each time step, $k$, the following steps were performed: (i) acquire a phase measurement, $\varphi_k$; (ii) compute the corresponding correction, $\Psi_k$, using the procedure from Sect.~\ref{sec:procedure}; (iii) move the phase screen by a step of $\delta$, simulating the wind-induced motion of the atmosphere in a $1/1500$\,s interval on an eight-meter-size telescope (a lateral shift of $a\times\delta$ is also applied, $a$ denoting a random number in the range $[-1:1])$; and (iv) apply the correction, $\Psi_{k-d}$, on the SLM with a gain of 0.5.

Figure~\ref{fig:wind_speed10} presents $\sigma_{OPD}$ for the standard deviation as seen by ZELDA in the OPD as a function of iterations for three different cases. While the starting standard deviation in the OPD is between 21 and 23.5\,nm RMS for all the configurations, there is a clear reduction in the wavefront error after fewer than ten iterations. However, the mean level reached for each curve that represents a filtered OPD is different. The highest threshold at 16.9\,nm RMS is reached with the two-frame delay configuration with the lowest filter at 44\,c/p (in red): The controllable modes are restricted by the low-pass filter. As a result, the contribution to wavefront error coming from uncontrolled high spatial frequencies, or fitting error, is higher. By increasing the filter frequency cutoff to 55\,c/p (in blue) while maintaining the two-frame delay, we reduced this fitting error and therefore improved the final correction quality to 15.6\,nm RMS. We finally considered the hypothetic case where the lag error could have been reduced to one frame (in green) with the filter at 44\,c/p. This lag reduction provides a clear improvement in the correction quality, which reduces to 14.6\,nm RMS, yielding a gain of 2.3\,nm RMS, on average, with respect to its two-frame lag counterpart. The final threshold in all three configurations is noticeably higher than what was achievable in Sect.~\ref{sec:static} due, in large part, to the fitting error of approximately 10\,nm. Furthermore, we used a loop gain of 0.5, which will undeniably limit the ultimate performances in a dynamic case. We did not identify the exact origin of the periodic pattern shown in Fig.~\ref{fig:wind_speed10}. Nevertheless, the fact it is identical for each realization when the same turbulence is used leads us to conclude that it is inherent to the phase screen rather than a result of our analysis.

\begin{figure}
    \centering
    \includegraphics[width=\columnwidth]{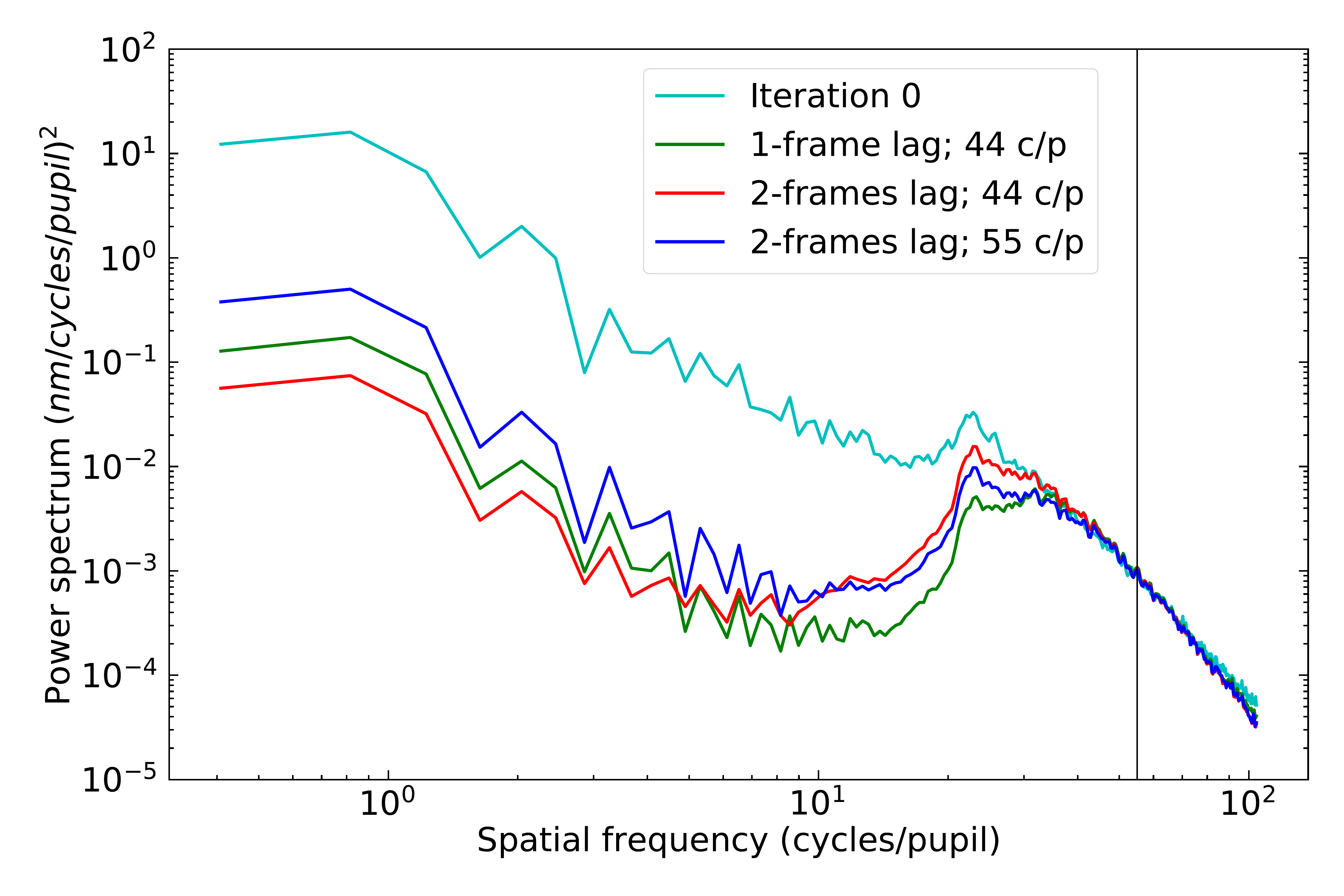}
    \caption{PSD of the OPD maps at simulated times of 0\,ms (cyan) and 40\,ms for the two-frame delay with $w$=55\,c/p (blue) and $w$=44\,c/p (red) and the one-frame delay with $w$=44\,c/p (green). The vertical black line represents the cutoff frequency at 55\,c/p.}
    \label{fig:power_spectrum}
\end{figure}

Figure~\ref{fig:power_spectrum} provides an additional description of the distribution of aberrations with spatial frequency in these three different cases after 40\,ms of simulated time, which is equivalent to 60 iterations. The plateaus reached in Fig.~\ref{fig:wind_speed10} show that other iteration values would yield the same power density spectrum (PSD)  behavior. The gain with respect to the initial phase step (in blue) is clear for the three PSD curves in the low- and mid-spatial frequencies, between 0 and 20\,c/p. Between 10 and 40\,c/p are the differences in terms of the OPD of Fig.~\ref{fig:wind_speed10}: For the two-frame delay curves, the stronger apodization of the filter at 44\,c/p translates into a higher PSD around the 20-40\,c/p area. Except for the low frequencies that are crippled by the bench turbulence, the best correction is provided by the reduced lag simulation (in green). The peak close to 20\,c/p corresponds to the cutoff frequency of the VLT/SAXO system that was simulated by the rotating phase screen.

In this experiment we performed two camera acquisitions for each phase measurement: one without the ZELDA mask to obtain a clear pupil image and one with the mask. This approach is not really suitable for an on-sky system as moving a mask at this cadence would be difficult. Furthermore, it would need to work solely based on the ZELDA image to maximize efficiency. Relying on a clear pupil measurement made at the beginning for the rest of the sequence has been investigated on MITHiC; unfortunately, it causes the correction to diverge, due to Fresnel propagation effects that modify the pupil intensity distribution, and biases the phase reconstruction. Based on the work of \citet{Vigan2019}, we expect this effect to not be present in VLT/SPHERE data. In Fig.~C.1  of that paper, the authors show the on-sky clear pupil image that they use for their correction sequence, which shows no sign of the turbulent phase. Although it is not directly reported in their work, they confirm that the clear pupil image is highly stable over time (private communication). However, MITHiC has not been designed to minimize these amplitude effects, and the test bed is very sensitive to misalignment or to phase errors along the optical axis because we scale down the eight-meter-class telescope turbulence on the 5 mm test bed pupil.

\subsection{Impact on coronagraphic images}
\label{sec:corono_results}

\begin{figure*}
    \centering
    \includegraphics[width=\textwidth]{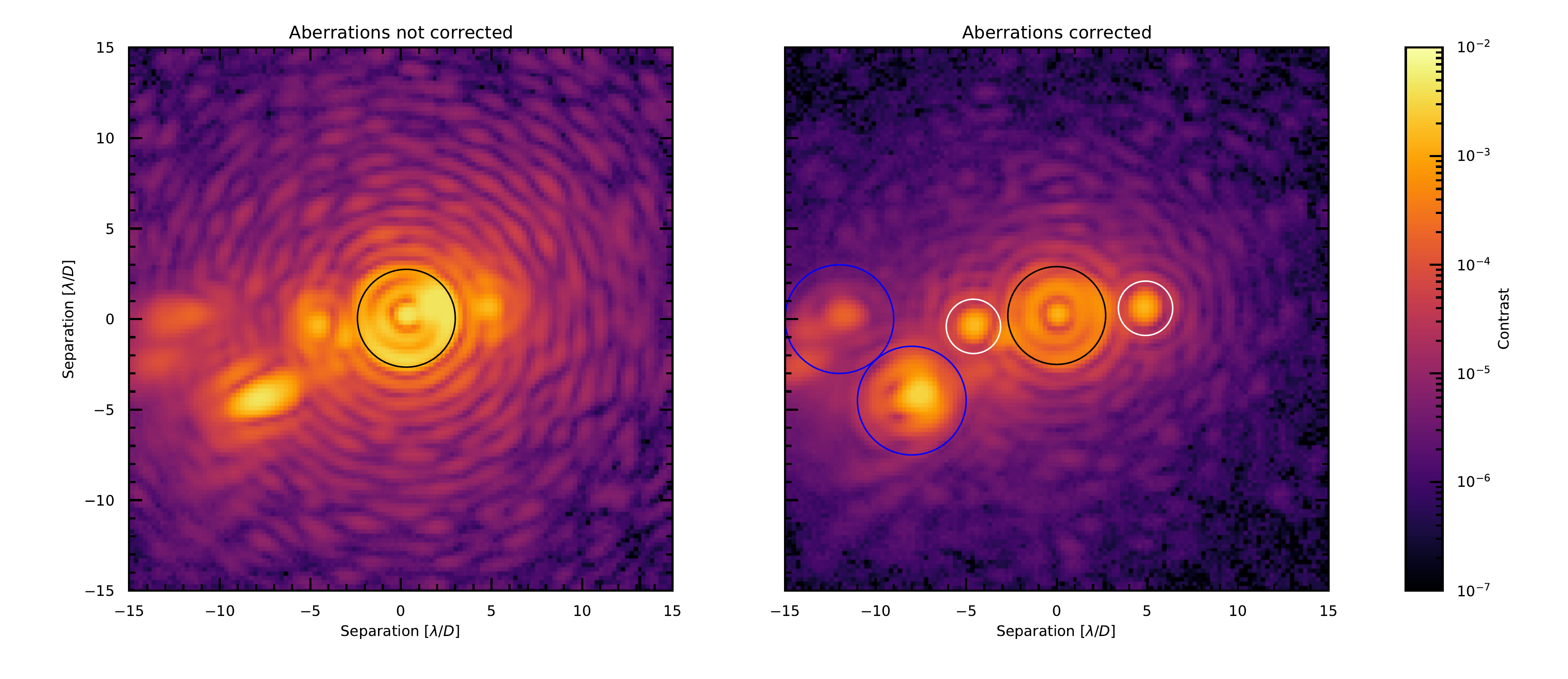}
    \caption{Coronagraphic images obtained before (left) and after (right) compensation of the bench aberrations. The compensation is performed with a low-pass filtering value of 125\,c/p. The CLC FPM has a radius of 2.7\,\lsd, the footprint of which is circled in black. The left part of the images is contaminated by several ghosts, circled in blue, which were introduced by the beam splitter located between the SLM and the coronagraphic FPM, which is used to send part of the beam to ZELDA. The speckles circled in white are created by ghost reflections in the cubic beam splitter in front of the SLM.}
    \label{fig:clc_images}
\end{figure*}

\begin{figure}
    \centering
    \includegraphics[width=\columnwidth]{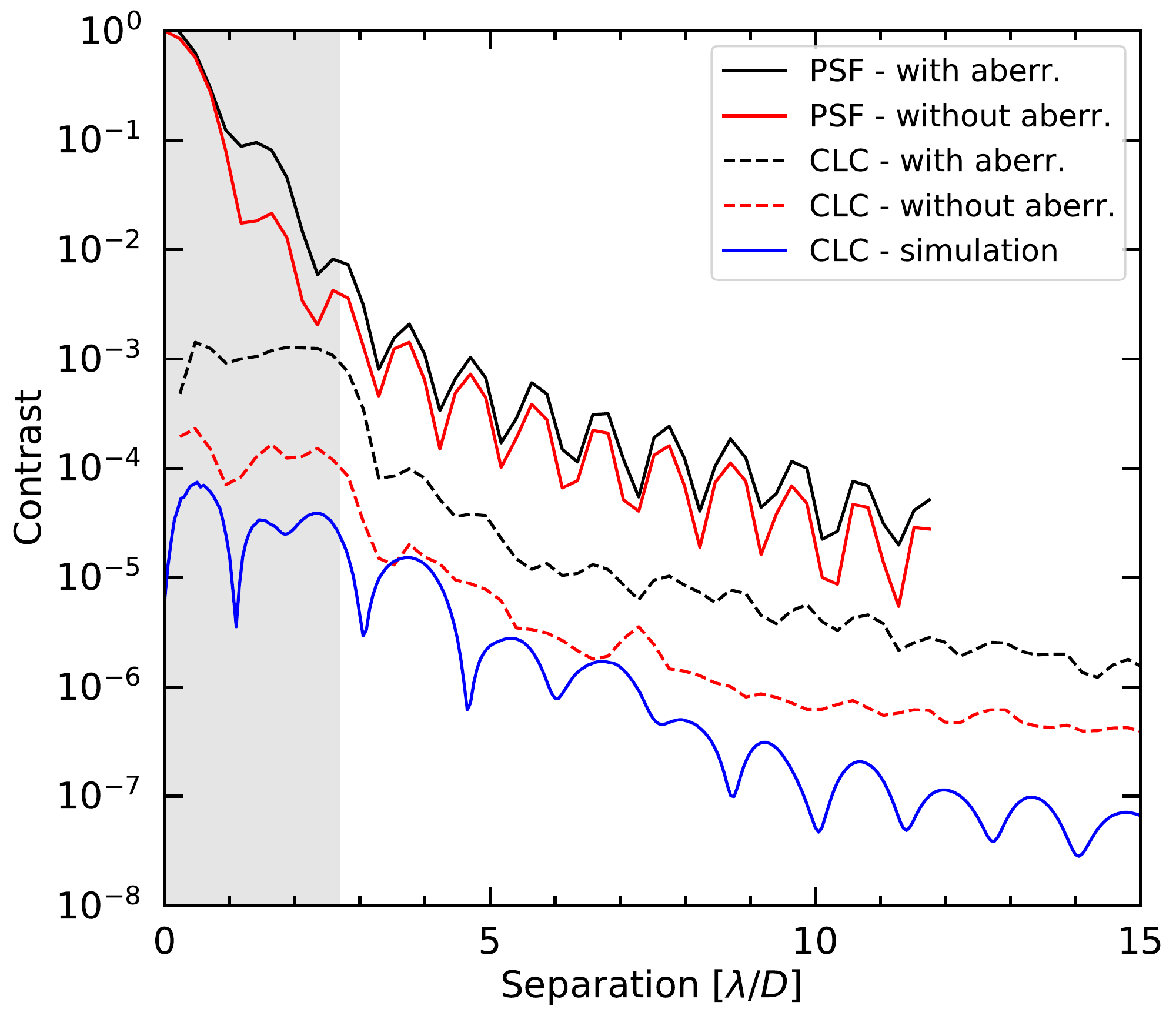}
    \caption{PSF and coronagraphic profiles before (black) and after (red) compensation of the bench aberrations. The profiles for the PSF and coronagraphic data are obtained with the azimuthal average and standard deviation of the images, respectively. The profiles are compared to the theoretical simulated profile (blue) obtained using as input an OPD map measurement with the bench aberration compensation prior to the coronagraphic measurement. The ghosts visible in Fig.~\ref{fig:clc_images} are masked when computing the profiles. The grayed out part of the plot corresponds to the area masked by the FPM in the coronagraphic images.}
    \label{fig:clc_profiles}
\end{figure}

The contrast performance of coronagraphs is crucially dependent on the quality of the incoming wavefront at the level of the FPM. Our bench calibration procedure allows for an accurate compensation for the MITHiC aberrations down to a few nanometer RMS. With such a high level of wavefront correction, MITHiC becomes diffraction-limited and should offer a significant attenuation of the diffraction with the CLC. While the CLC provides a moderate contrast \citep[compared with the apodized-pupil Lyot coronagraph;][]{Soummer2005}, and a moderate inner working angle \citep[compared with the vortex coronagraph;][]{Mawet2005}, it remains a user-friendly coronagraph for simple coronagraphic imaging in a monochromatic light system that works with a circular pupil, such as MITHiC.

 We performed coronagraphic measurements in the initial configuration of MITHiC (i.e., without aberration compensation and after five iterations of the compensation procedure with $w=$125\,c/p), which means that most of the controllable spatial frequencies with the SLM are corrected. Figure~\ref{fig:clc_images} shows the coronagraphic images before and after the aberration compensation. Due to the use of a beam splitter to send part of the beam to ZELDA, the left part of the images is dominated by relatively bright ghosts, which are circled in blue. The two symmetric ghosts circled in white on either side of the PSF are induced by the other beam splitter located in the collimated beam that hits the SLM. Ignoring the ghosts, the residual aberration correction results in a substantial improvement of the diffraction attenuation of the coronagraph, with a visible decrease in the amount of light scattered in speckles all over the field of view. Before calibration, the focal plane shows diffraction rings, speckles, and pinned speckles \citep{Sivaramakrishnan2002} up to 15\,\lsd. After calibration, the brightness of the speckles and residual diffraction appears largely attenuated, indicating an improvement in the CLC performance. Apart from the four ghost images, some irregular speckle patterns are visible at larger separations in the coronagraphic images. This indicates the presence of residual high-spatial-frequency aberrations that could originate from NCPAs between the ZELDA arm and the coronagraph arm. 

The gain in contrast performance of the CLC is confirmed by the PSF and coronagraphic image profiles in Fig.~\ref{fig:clc_profiles}, which show a clear distinction before and after residual aberration compensation. This is particularly visible for the coronagraphic profiles, with a mean gain of a factor of five between 3 and 15\,\lsd. We compare the aberration-corrected coronagraphic profile to a CLC simulation that takes a residual OPD map measured with ZELDA prior to the coronagraphic measurement as input. Between 3 and 7\,\lsd, our measurements are clearly at the theoretical limit of the CLC implemented on MITHiC. At larger separations, there is a slow departure between the theoretical and experimental profiles, which can be explained by some residual aberrations and by the readout noise floor of the camera above $\sim$12\,\lsd.

\section{Conclusion}
\label{sec:conclusion}

The wavefront quality is a driving parameter in the performance of high-contrast imaging systems that use coronagraphy to attenuate telescope diffraction on an observed star image. ZELDA has already been demonstrated to be a powerful secondary WFS that can calibrate aberrations unseen by an ExAO system.

In the present work, we demonstrated the ability of ZELDA to work in a closed loop with an SLM to perform a second-stage AO correction, controlling a very large number of degrees of freedom.
Thanks to the use of a high-density SLM, the MITHiC test bed offers tens of thousands of DM-like actuators for wavefront correction. While these numbers appear large by today's standards, they are in line with the expectations for high-contrast imaging of Earth-like planets with future ground-based ELTs. Rather than building a large interaction matrix, we instead implemented a direct wavefront correction based on the OPD maps obtained with ZELDA and complemented by a simple and fast distortion calibration procedure. Our results demonstrate that this procedure is very efficient at compensating for the static phase patterns of the bench and remaining stable once a minimum has been reached. With the current procedure, the correction for the wavefront errors in a static case reaches a level of $\sim$2.6\,nm\,RMS. 

The pupil distortion calibration is a crucial step. While such effects are naturally taken into account in systems relying on a classical interaction matrix, this is not the case here. We instead used a two-frame calibration where Gaussian reference spots are introduced on the SLM. Based on their known input position, we were able to easily calibrate the distortions in the system and subsequently correct for them when applying the phase corrections on the SLM. Without this calibration, the wavefront errors quickly diverge in the phase compensation procedure due to the apparition of high-spatial-frequency ripples induced by the distortion.

We also demonstrate a dynamic ``fast-ZELDA'' mode with a more dynamic case where the sensor is used to compensate for the residuals of a VLT/SPHERE-like ExAO system. This setup acts as a second-stage correction that could possibly be used in very demanding science cases. For this demonstration, we used the MITHiC phase screen, which allows for the simulation of VLT/SPHERE-like turbulence residuals. We simulated a system running at 1.5\,kHz with a delay of one or two frames in the correction of the residuals. The results show an efficient wavefront correction that is stable over a thousand iterations. While the simulated residual aberrations present a wavefront error of around 20-25\,nm\,RMS, the closed-loop correction, dominated by the uncontrollable fitting errors and the bench internal turbulence, reduces this to approximately 15\,nm\,RMS.

Our setup on MITHiC presents some shortcomings. Firstly, we work with an SLM in monochromatic light. From the WFS point of view, similar results could certainly be obtained with a reasonable broadband since ZELDA has been shown to accommodate bandwidths of 20\% well \citep{N'Diaye2013, N'Diaye2016}. However, the SLM is chromatic and cannot be used in a broadband setup. Broadband demonstration would require the implementation of a classical high-density DM. Secondly, our measurement procedure requires two camera acquisitions for one OPD measurement: with and without the ZELDA mask. This would be an issue when running a system at several kilohertz. Further studies need to be performed on the stability of the amplitude in the pupil, as well as on whether or not taking the acquisition without the mask is mandatory. Nevertheless, we could foresee a fast synchronized tip and tilt mirror that would temporally switch the beam from one configuration to the other or a setup with a beam splitter to perform the two simultaneous measurements. Such an idea has been developed with the vector ZWFS, which splits the beam into the two circular polarizations, which face respectively a $+\pi/2$ and $-\pi/2$ Zernike phase mask. \citep{2020PASP..132d5002D}.

The coronagraphic measurements with the CLC on MITHiC confirm the extreme efficiency of our residual aberration compensation. Despite its simple design, the CLC becomes an efficient solution for diffraction attenuation in monochromatic light and with an unobstructed circular pupil. Combined with ZELDA for the fine phase correction, this setup is now the baseline for the forthcoming studies with MITHiC, which are centered on laboratory investigations of fiber injection procedures for the High-Resolution Imaging and Spectroscopy of Exoplanets (HiRISE) project \citep{Vigan2018,Otten2020}. These investigations will advance wavefront control techniques for future exoplanet imagers.

\begin{acknowledgements}
    AV acknowledges support from R\'egion Provence-Alpes-C\^ote d'Azur, under grant agreement 2014-02976, for the ASOREX project. AV, ME, GO and ML acknowledge funding from the European Research Council (ERC) under the European Union's Horizon 2020 research and innovation programme (grant agreement No.~757561).
    
    ER and JH have benefited from the support of the A*MIDEX university foundation while following the Erasmus Mundus Europhotonics Master program of the European Union.
\end{acknowledgements}

\bibliographystyle{aa}
\bibliography{biblio}

\appendix

\section{Detailed description of MITHiC}
\label{sec:mithic_detailed}

This section provides a more in-depth description of MITHiC. Three main parts of the test bed (see Fig.~\ref{fig:mithic}) are described in dedicated subsections: the telescope and ExAO in Sect.~\ref{sec:telescope}, the second-stage AO part in Sect.~\ref{sec:zelda}, and the coronagraph part in Sect.~\ref{sec:corono}. In addition, we briefly describe the bench control in Sect.~\ref{sec:control}.

Generally speaking, all the lenses on the bench are two-inch off-the-shelf achromatic doublets coated for visible and near-infrared applications, and one-inch high-quality beam splitters. The optics are oversized compared to the beam size, which is usually smaller than 5\,mm. Using large optics with high optical quality allows us to reduce the amount of aberrations, with a beam only hitting at a small fraction of the high-quality optic, and to limit any vignetting effect of the diffracted beams.

\subsection{Telescope and ExAO}
\label{sec:telescope}

The telescope part goes from the light source to a dedicated phase screen and simulates a telescope observing an unresolved star. The light source is a fiber-coupled superluminescent diode, which has a central wavelength $\lambda = 670.7$\,nm and a spectral bandwidth of 10\,nm. The controller of the light source is connected to an Arduino Uno board that can switch the source on or off on commands received from the MITHiC control computer. The light is linearly polarized to match the SLM requirement, using a combination of a linear polarizer and a quarter-wave plate. The light is then injected into a polarization-maintaining fiber. The output end of the fiber is mounted into a fiber-optic positioner with six degrees of freedom and serves as an unresolved point source at the entrance of the bench. The point source directly illuminates the circular, unobstructed entrance pupil. By selecting the core part of the Gaussian beam, we approximate a flat wavefront with a homogeneous illumination within the pupil. This approximation is valid given the small numerical aperture of the fiber (NA$\simeq$0.12), the large distance between the output of the fiber and the pupil ($\sim$15\,cm), and the small pupil of diameter 4.96\,mm.

A phase screen is then used to simulate various types of aberration patterns that are typically encountered in large optical telescopes. The phase screen was manufactured by SILIOS Technologies, based on specifications for MITHiC \citep{Vigan2016SPIE}. It consists of a 100-millimeter-diameter fused silica substrate in which phase patterns are engraved using cumulative etching technology. This technology allows for building multilevel stair-like topologies directly into fused silica substrates \citep{Caillat2014}. It is based on successive masking photolithography and reactive ion etching steps (i.e., successive etching steps through resin masks). The whole phase pattern is thus engraved at the same time, resulting in a very high uniformity over the full engraved area. The MITHiC phase screen includes a dozen static patterns representing standard NCPAs, segmented pupils with piston, tip and tilt aberrations, turbulence, or low-wind effects as visible on VLT/SPHERE \citep{Sauvage2015}. In addition, the phase screen includes two continuous rings of simulated residual turbulence for VLT/SPHERE and ELT/HARMONI. The phase screen is mounted in a custom structure designed by SILIOS, which includes a stepper motor that enables a fast rotation of the phase screen to select specific phase patterns or continuous motion for the simulated residual turbulence. The phase screen is mounted on a precision motorized linear stage to move it in translation with respect to the optical axis.

\subsection{Second-stage AO}
\label{sec:zelda}

The second part corresponds to the second-stage AO part, which uses an SLM for the wavefront correction and a ZELDA WFS.

The SLM from Hamamatsu acts as a DM. It is made of liquid crystal pixels that can each induce independent phase shifts and offers a fill factor of 98\%. There are 274 SLM pixels across the pupil diameter, $D_{\mathrm{SLM}}=$5.5\,mm, resulting in almost 60\,000 independent resolution elements in the pupil. The SLM takes an eight-bit image as an input, allowing for 256 evenly spaced phase values for each pixel. At $\lambda = 670.7\,nm$, a $2\pi$ phase shift yields a pixel value of $N_{steps}$ of 254. The resolution, $\Delta_{\mathrm{SLM}}$, in terms of OPD is therefore $\Delta_{\mathrm{SLM}} = \lambda / N_{\mathrm{steps}} \approx 2.64$\,nm. A low limit of the wavefront error RMS value achievable by such a device is given by the quantization noise: $\sqrt{\Delta_{\mathrm{SLM}}^2/12}=0.76$\,nm\,RMS. The SLM is connected to a dedicated computer running Linux, on which a Python server runs and displays the eight-bit images sent by the MITHiC control computer over the local network. The SLM is used in reflection, so it is preceded by a high-quality beam splitter. While this configuration induces a loss of 75\% of the flux because of the double pass, it allows us to keep the SLM perpendicular to the optical beam.

Following the SLM, a beam splitter sends half of the light toward ZELDA for wavefront sensing. The ZELDA phase mask is located in an F/90 focal plane. It was manufactured in fused silica by SILIOS with photolithography, following the process described in \citet{N'Diaye2010}. Its depth is 350\,nm, corresponding to nearly $\lambda/4$ in optical depth at 670.7\,nm, and it has a diameter of 64\,\mic, corresponding to a relative size of 1\lsd \citep{N'Diaye2013}. The phase mask is mounted on a three-axis mount motorized with high-resolution linear actuators, which enable its positioning with 1\,\mic accuracy.

In the next pupil plane, a Hamamatsu ORCA-Flash4.0 V3 CMOS camera is used to reimage the pupil. In this plane, the pupil image has a diameter $D_{WFS}=$2.75\,mm, which corresponds to 418 pixels on the detector, and provides more than 130,000 single-resolution elements for wavefront sensing.

The ZELDA WFS arm provides wavefront sensing in a plane close to the coronagraphic FPM, which removes most of the NCPA in traditional benches or instruments. The only differential optical element here is a cubic beam splitter. The remaining NCPA would have to be calibrated to enhance the wavefront correction in the coronagraphic focal plane. This can be done, for example, by placing another ZELDA mask instead of the coronagraph FPM, which has not been used in this paper.

\subsection{Coronagraphic imaging}
\label{sec:corono}

The last part of the bench corresponds to the coronagraph and associated science imaging channel. The bench includes a focal plane at F/90 that is immediately followed by a pupil plane, which enables the implementation of all coronagraphs based on the combination of an FPM and a Lyot stop \citep{Lyot1932}. Although there is no available pupil plane upstream from the FPM, the beam is collimated after the SLM, which enables the insertion of an apodizer if necessary.

The bench was originally designed to test the Roddier \& Roddier phase mask coronagraph, both in its original form \citep{Roddier1997} and with an apodizer \citep{Soummer2003a, N'Diaye2012b}. While these coronagraphs can in theory provide extremely deep raw contrasts, they are also very sensitive to aberrations, vibrations, centering, etc. Moreover, SLMs are notably affected by fill factor issues \citep{Strauss2016, Ronzitti2012}, which cause some of the incoming wavefront to not ``see'' the liquid crystal pixels; as a result, they are not corrected. This results in a partially uncorrected PSF at the level of the FPM, which affects the final performance, in particular at very small inner working angles ($\sim$\lsd) where the Roddier \& Roddier coronagraph is expected to be particularly interesting.

To alleviate some of the performance issues at small inner working angles, we recently implemented a CLC based on an opaque FPM and an associated Lyot stop. The diameters of the FPM and Lyot stop were numerically optimized to maximize the raw image contrast for separations ranging between 5 and 10\lsd over a grid of sizes for both components, leading us to diameters of 5.38\lsd for the FPM and 0.77$D$ for the Lyot stop. This setup will, in theory, allow us to reach a mean contrast of $2 \times 10^{-6}$ in the 5--10\lsd range with 10\,nm\,RMS of static aberrations. The FPM was manufactured by OPTIMASK using a chromium deposit on a BK7 substrate, providing an optical density of more than four at our working wavelength. The Lyot stop was manufactured by STEEC using laser cutting in a 100\,\mic stainless steel plate. Similarly to the ZELDA phase mask, the coronagraphic FPM is mounted on a controllable three-axis stage. The Lyot stop is simply mounted on a three-axis stage that is manually movable with micrometric screws.

The science imaging camera is a Coolsnap HQ2 CCD camera from Teledyne Photometric that provides a fast readout with low noise. The camera can be used to image the coronagraphic signal either in the focal plane or in the pupil plane with an optional lens.

\subsection{Control computer}
\label{sec:control}

All the active elements on the bench are controlled by a single computer. We have developed dedicated control software written entirely in Python, which is interfaced with all controllable elements: the light source controller, the phase screen translation and rotation stages, the three-axis stages for the ZELDA and coronagraphic FPMs, and the ZELDA and science cameras. On the bench, we also have USB-controlled humidity and temperature sensors from Thorlabs, as well as an Arduino Nano 33 BLE board with an embedded accelerometer working at 800\,Hz to monitor vibrations.

The controller of the source is connected to an Arduino board running custom software that is controlled with serial-over-USB using \texttt{pySerial}\footnote{\url{https://pythonhosted.org/pyserial/}}. The control only offers an on--off switch of the source, so the power of the source is set to a constant level that satisfies the various uses of the bench.

All the precision linear stages from Physik Instrumente (PI) are controlled using the official \texttt{PIPython}\footnote{\url{https://github.com/git-anonymous/PIPython}} module, which wraps the PI general command set (GCS) and provides methods to connect to the devices. The stepper for the rotation of the phase screen is controlled with \texttt{pySerial} and is based on the official serial commands from the manufacturer of the motor. The Thorlabs elements are controlled using custom python modules that directly interface with the drivers using the Python built-in \texttt{ctypes}\footnote{\url{https://docs.python.org/3/library/ctypes.html}} module.

The cameras are controlled with a custom wrapper that supports different types of cameras. The Hamamatsu camera for ZELDA is supported using a custom module that interfaces with the official driver from Hamamatsu named DCAM, again using \texttt{ctypes}. For the science camera, we use the official \texttt{PVCAM}\footnote{\url{https://github.com/Photometrics/PyVCAM}} module from Teledyne Photometric.

Finally, the SLM is controlled through a client and server system, with the client running on the MITHiC control computer and the server running on the dedicated SLM computer. The client directly sends eight-bit encoded images to be displayed on the SLM, and the server displays these images on a full-screen window using the \texttt{Qt} graphical user interface toolkit\footnote{\url{https://www.qt.io/}} and its official Python implementation, \texttt{PySide2}\footnote{\url{https://wiki.qt.io/Qt_for_Python}}.

The advantage of the full Python implementation of the test bed control is the ability to easily script repetitive or long procedures, such as ZELDA and coronagraph FPM centering, coronagraphic data acquisition sequences, calibrations, etc. Covered by an enclosure, the bench is also fully operable remotely over the network using remote desktop functionalities, which limits sources of vibration and temperature variations, air movements, and dust accumulation in the MITHiC room.

\end{document}